\documentclass[12pt]{article}

\usepackage{scicite}

\usepackage{times}

\usepackage{graphicx}
\usepackage{float}
\usepackage{amsmath}
\usepackage{amssymb}
\usepackage{upgreek}

\makeatletter
\renewenvironment{abstract}{%
    \if@twocolumn
      \section*{\abstractname}%
    \else 
      \begin{center}%
        {\bfseries \Large\abstractname\vspace{\z@}}
      \end{center}%
      \quotation
    \fi}
    {\if@twocolumn\else\endquotation\fi}
\makeatother

\newcounter{lastnote}

\title{A Robotic Model of Hippocampal Reverse Replay for Reinforcement Learning}

\author
{Matthew T. Whelan,$^{1,2}$ Tony J. Prescott,$^{1,2}$ Eleni Vasilaki$^{1,2,*}$\\
\\
\normalsize{$^{1}$Department of Computer Science, The University of Sheffield, Sheffield, UK}\\
\normalsize{$^{2}$Sheffield Robotics, Sheffield, UK}\\
\normalsize{$^{*}$Corresponding author: e.vasilaki@sheffield.ac.uk}
}

\date{}

\begin{document} 


\baselineskip24pt


\maketitle


\begin{abstract}
\normalsize{Hippocampal reverse replay is thought to contribute to learning, and particularly reinforcement learning, in animals. We present a computational model of learning in the hippocampus that builds on a previous model of the hippocampal-striatal network viewed as implementing a three-factor reinforcement learning rule. To augment this model with hippocampal reverse replay, a novel policy gradient learning rule is derived that associates place cell activity with responses in cells representing actions. This new model is evaluated using a simulated robot spatial navigation task inspired by the Morris water maze. Results show that reverse replay can accelerate learning from reinforcement, whilst improving stability and robustness over multiple trials. As implied by the neurobiological data, our study implies that reverse replay can make a significant positive contribution to reinforcement learning, although learning that is less efficient and less stable is possible in its absence. We conclude that reverse replay may enhance reinforcement learning in the mammalian hippocampal-striatal system rather than provide its core mechanism.}
\end{abstract}


\newpage
\section{Introduction}

Many of the challenges in the development of effective and adaptable robots can be posed as reinforcement learning (RL) problems; consequently there has been no shortage of attempts to apply RL methods to robotics \cite{kober2013reinforcement, sutton2018reinforcement}. However, robotics also poses significant challenges for RL systems. These include factors such as continuous state and action spaces, real-time and end-to-end learning, reward signalling, behavioural traps, computational efficiency, limited training examples, non-episodic resetting, and lack of convergence due to non-stationary environments \cite{kober2013reinforcement, kuutti2020survey, zhu2020ingredients}. 

Much of RL theory has been inspired by early behavioural studies in animals \cite{sutton2018reinforcement}, and for good reason, since biology has found many of the solutions to the control problems we are searching for in our engineered systems. As such, with continued developments in biology, and particularly in neuroscience, it would be wise to continue transferring insights from biology into robotics \cite{prescott2018living}. Yet equally important is its inverse, the use our computational and robotic models to inform our understanding of the biology \cite{webb2001can, mitchinson2011biomimetic}. Robots offer a valuable real-world testing opportunity to validate computational neuroscience models \cite{sheynikhovich2009there, jauffret2015grid, prescott2019memory}. 

Though the neurobiology of RL has largely centred on the role of dopamine as a reward-prediction error signal \cite{schultz1998predictive, redgrave2017phasic}, there are still questions surrounding how brain regions might coordinate with dopamine release for effective learning. Behavioural timescales evolve over seconds, perhaps longer, whilst the timescales for synaptic plasticity in mechanisms such as spike-timing dependent plasticity (STDP) evolve over milliseconds \cite{bi1998synaptic} -- how does the nervous system bridge these time differentials so that rewarded behaviour is reinforced at the level of synaptic plasticities?

One recent hypothesis offering an explanation to this problem has been in three-factor learning rules \cite{vasilaki2009spike,richmond2011democratic,fremaux2016neuromodulated,gerstner2018eligibility}. In the three-factor learning rule hypothesis, learning at synapses occurs only in the presence of a third factor, with the first and second factors being the typical pre- and post-synaptic activities. This can be stated in a general form as follows,
\begin{equation}
\label{eqtn:general_form_three_factor_learning_rule}
    \frac{d}{dt}w_{ij} = \eta f(x_j)g(y_i)M(t)
\end{equation}
where $\eta$ is the learning rate, $x_j$ represents a pre-synaptic neuron with index $j$, $y_i$ a post-synaptic neuron with index $i$, and $f(\cdot)$ and $g(\cdot)$ being functions mapping respectively the pre- and post-synaptic neuron activities. $M(t)$ represents the third factor, which here is not specific to the neuron indices $i$ and $j$ and is therefore a global term. This third factor is speculated to represent a neuromodulatory signal, which in this case is best thought of as dopamine, or more generally as a reward signal. Equation \ref{eqtn:general_form_three_factor_learning_rule} appears to possess the problem stated above, of how learning can occur for neurons that were co-active prior to the introduction of the third factor. To solve this, a synaptic-specific eligibility trace is introduced, which is a time-decaying form of the pre- and post-synaptic activities \cite{gerstner2018eligibility},
\begin{equation}
    \label{eqtn:three-factor learning rule with elig}
    \begin{split}
    &\frac{d}{dt}e_{ij} = - \frac{e_{ij}}{\uptau_e} + \eta f(x_j)g(y_i) \\
    &\frac{d}{dt}w_{ij} = e_{ij}M(t)    
    \end{split}
\end{equation}
The eligibility trace time constant, $\uptau_e$, modulates how far back in time two neurons were co-active in order for learning to occur -- the larger $\uptau_e$ is, the more of the behavioural time history will be learned and therefore reinforced. To effectively learn behavioural sequences over the time course of seconds $\uptau_e$ is set to be in the range of a few seconds \cite{gerstner2018eligibility}. Work conducted by Vasilaki et al. \cite{vasilaki2009spike} successfully applied such a learning mechanism in a spiking network model for a simulated agent learning to navigate in a Morris water maze task \cite{vasilaki2009spike}, in which they used a value of 5s for $\uptau_e$, a value that was optimised to that specific setting.

\textit{Hippocampal replay} however suggests an alternative approach, building on the three-factor learning rule but avoiding the need for synapse-specific eligibility traces. Hippocampal replay was originally shown in rodents as the reactivation during sleep states of hippocampal place cells that were active during a prior awake behavioural episode \cite{wilson1994reactivation, skaggs1996replay}. During replay events, the place cells retain the temporal ordering experienced during the awake behavioural state, but do so on a compressed timescale -- replays typically replay cell activities over the course of a few tenths of a second, as opposed to the few seconds it took during awake behaviour. Furthermore, experimental results presented later to these original results showed that replays can occur in the \textit{reverse} direction too, and that these \textit{reverse replays} occurred when the rodent had just reached a reward location \cite{foster2006reverse, diba2007forward}. Interestingly, these replays would repeat the rodent's immediate behavioural sequence that had led up to the reward, which led Foster and Wilson \cite{foster2006reverse} to speculate that hippocampal reverse replays, coupled with phasic dopamine release, might be such a mechanism to reinforce behavioural sequences leading to rewards. 

Whilst it has been well established that hippocampal neurons project to the nucleus accumbens \cite{humphries2010ventral}, the proposal that reverse replays may play an important role in RL has since received further support. For instance, there are experimental results showing that reverse replays often co-occur with replays of the ventral striatum \cite{pennartz2004ventral} as well as there being increased activity in the ventral tegmental area during awake replays \cite{gomperts2015vta}, which is an important region for dopamine release. Furthermore, rewards have been shown to modulate the frequency with which reverse replays occur, such that increased rewards promotes more reverse replays, whilst decreased rewards suppresses reverse replays \cite{ambrose2016reverse}. 


To help better understand the role of hippocampal reverse replays in the RL process, we present here a neural RL network model that has been augmented with a hippocampal CA3 inspired network capable of producing reverse replays. The network has been implemented on a simulation of the biomimetic robot MiRo-e \cite{mitchinson2016miro} to show its effectiveness in a robotic setting. The RL model is an adapted hippocampal-striatal inspired spiking network by \cite{vasilaki2009spike} derived using a policy gradient method, but modified here for continuous-rate valued neurons -- this modification leads to a novel learning rule, though similar to previous learning rules \cite{williams1992simple}, for this particular network architecture. The hippocampal reverse replay network meanwhile is taken from Whelan et al. \cite{whelan2020fast}, who implemented the network on the same MiRo robot, itself based on earlier work by Haga and Fukai \cite{haga2018recurrent} and Pang and Fairhall \cite{pang2019fast}. We demonstrate in robotic simulations that activity replay improves the stability and robustness of the RL algorithm by providing an additional signal to the eligibility trace.

\section{Methodology}
\subsection{MiRo Robot and the Testing Environment}
We implemented the model using a simulation of the biomimetic robot MiRo-e. The MiRo robot is a commercially available biomimetic robot developed by Consequential Robotics Ltd in partnership with the University of Sheffield. MiRo’s physical design and control system architecture are inspired by biology, psychology and neuroscience \cite{mitchinson2016miro} making it a useful platform for embedded testing of brain-inspired models of perception, memory and learning \cite{ling2019obstacle}. For mobility it is differentially driven, whilst for sensing we make use of its front facing sonar for the detection of approaching walls and objects. The Gazebo physics engine is used to perform simulations, where we take advantage of the readily available open-arena (Figure \ref{fig:Miro_with_place_action_cells}C). The simulator uses the Kinetic Kame distribution of the Robot Operating System (ROS). Full specifications for the MiRo robot, including instructions for simulator setup, can be found on the MiRo documentation web page \cite{consequentialroboticsdocs}.

\subsection{Network Architecture}
\begin{figure}[h!]
    \centering
    \includegraphics[width=0.75\linewidth]{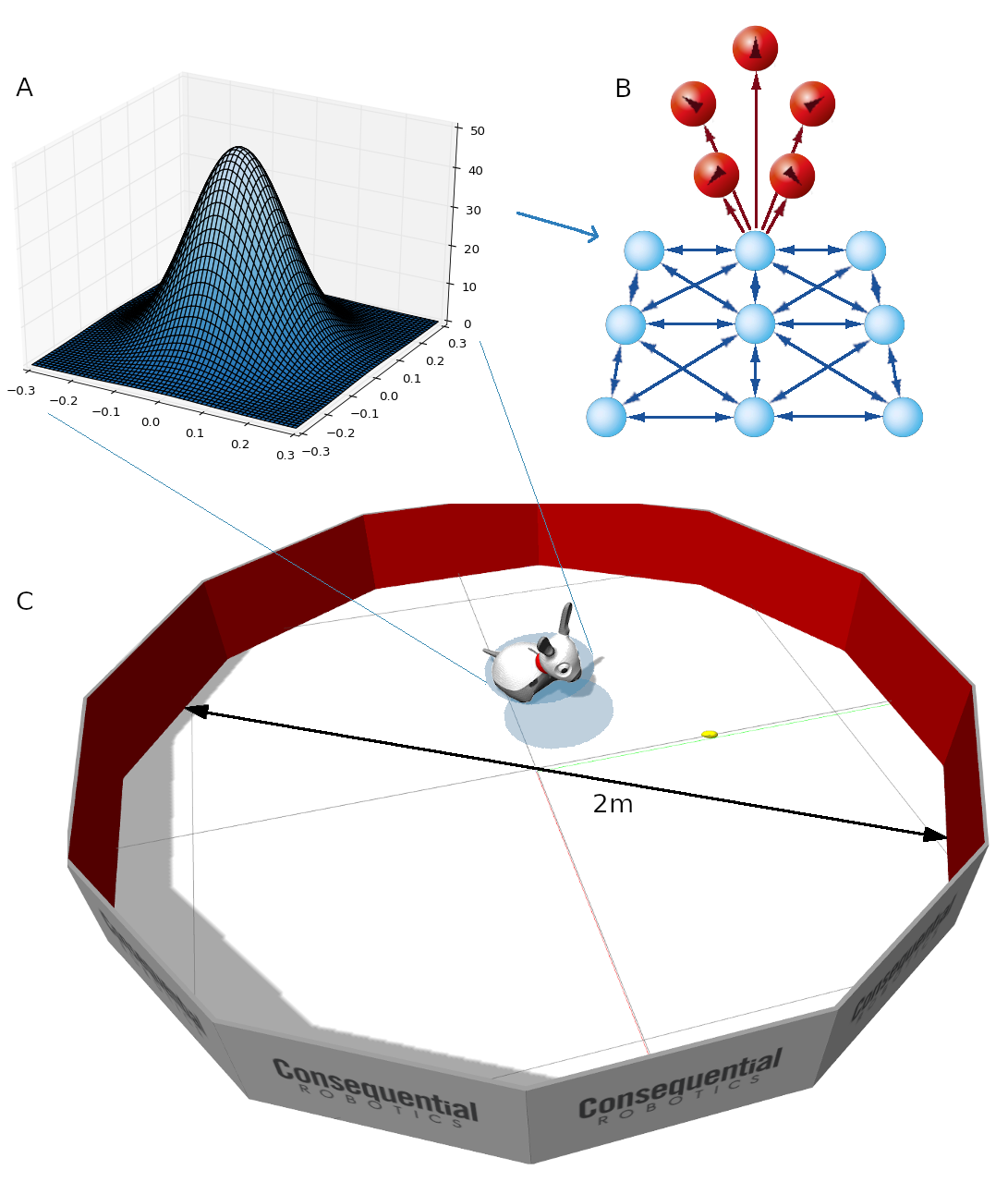}
    \caption{The testing environment, showing the simulated MiRo robot in a circular arena. A) Place fields are spread evenly across the environment, with some overlap, and place cell rates are determined by the normally distributed input computed as a function of MiRo's distance from the place field's centre. B) Place cells (blue, bottom set of neurons) are bidirectionally connected to their eight nearest neighbours. These synapses have no long-term plasticity, but do have short-term plasticity. Each place cell connects feedforwardly via long-term plastic synapses to a network of action cells (red, top set of neurons). In total there are 100 place cells and 72 action cells.}
    \label{fig:Miro_with_place_action_cells}
\end{figure}

The network is composed of a layer of 100 bidirectionally connected \textit{place cells}, which connects feedforwardly to a layer of 72 \textit{action cells} via a weight matrix of size 100$\times$72 (Figure \ref{fig:Miro_with_place_action_cells}B). In this model, activity in each of the place cells is set to encode a specific location in the environment \cite{okeefe1971hippocampus,okeefe1976place}. Place cell activities are generated heuristically using two dimensional normal distributions of activity inputs, determined as a function of MiRo's position from each place field's centre point (Figure \ref{fig:Miro_with_place_action_cells}A). This is similar to other approaches of place cell activity generation \cite{vasilaki2009spike,haga2018recurrent}. The action cells are driven by the place cells, with each action cell encoding for a specific with 5 degree increments, thus 72 action cells encode 360 degrees of possible heading directions. By computing a population vector of the action cell activities, these discreet heading directions can be transformed into continuous headings. For simplicity, MiRo's forward velocity is kept constant at 0.2m/s. We now describe the details of the network in full.

\subsubsection{Hippocampal Place Cells}
The network model of place cells represents a simplified hippocampal CA3 network that is capable of generating reverse replays of recent place cell sequence trajectories. This model of reverse replays was first presented in \cite{whelan2020fast}, but with one minor modification. Whereas the reverse replay model in \cite{whelan2020fast} has a global inhibitory term acting on all place cells, here the place cells have those inhibitory inputs removed from their dynamics. This modification does not affect the ability of the network to produce reverse replays, which is shown in the Supplementary Material, where reverse replays both with and without global inhibition are compared.


In more detail, the place cells consist of a network of 100 neurons, each of which is bidirectionally connected to its eight nearest neighbours as determined by the positioning of their place fields. Hence, place cells with neighbouring place fields are bidirectionally connected to one another (Figure \ref{fig:Miro_with_place_action_cells}B), whereas place cells whose place fields are further than one place field apart are not. In this manner, the connectivity of the network represents a map of the environment. This network approach is similar to the network approach taken by Haga and Fukai \cite{haga2018recurrent} in their model of reverse replay, except their weights are plastic whilst we keep ours static. The static weights for each cell, represented by $w_{jk}^{place}$ indicating the weight projecting from neuron $k$ onto neuron $j$, are all set to 1, with no cells self-projecting to themselves. Figure \ref{fig:Miro_with_place_action_cells}B displays the full connectivity schema for the bidirectionally connected place cell network.

The rate for each place cell neuron, represented by $x_j$, is given as a linearly rectified rate with upper and lower bounds,

\begin{equation}
    x_j = \begin{cases}
    0 &\text{if $x_j^{'} < 0$}\\
    100 &\text{if $x_j^{'} > 100$}\\
    x_j^{'} &\text{otherwise.}
    \end{cases}
\end{equation}
 The variable  $x_j^{'}$ is defined as, 
\begin{equation*}
    x_j^{'} = \alpha \left(I_j - \epsilon \right)
\end{equation*}

\noindent where $\alpha$, $\epsilon$ are constants and $I_j$ is the cell's activity, which evolves according to time decaying first order dynamics,
\begin{equation}
    \label{eqtn:place_cell_dynamics}
    \uptau_I\frac{d}{dt}I_j = -I_j + \psi_jI_j^{syn} + I_j^{place} 
\end{equation}
where $\uptau_I$ is the time constant, $I_j^{syn}$ is the synaptic inputs from the cell's neighbouring neurons, and $I_j^{place}$ is the place specific input calculated as per a normal distribution of MiRo's position from the place field's centre point. $\psi_j$ represents the place cell's \textit{intrinsic plasticity}, detailed further below.

Each place cell has associated with it a place field in the environment defined by its centre point and width, with place fields distributed evenly across the environment (100 in total). As stated, the place specific input, $I_j^{place}$, is computed from a two-dimensional normal distribution determined by MiRo's distance from the place field's centre point,
\begin{equation}
    \label{eqtn:place_cell_input}
    I^{place}_j = I^{p}_{max} \text{exp}\left[- \frac{(x_{MiRo}^c - x_j^c)^2 + (y_{MiRo}^c - y_j^c)^2}{2d^2} \right]
\end{equation}
where $I^{p}_{max}$ determines the max value for the place cell input. $(x_{MiRo}^c, y_{MiRo}^c)$ represents MiRo's $(x, y)$ coordinate position in the environment, whilst $(x_j^c, y_j^c)$ is the location of the place field's centre point. The term $d$ in the denominator is a constant that determines the width of the place field.

The synaptic inputs, $I_j^{syn}$, are computed as a sum over neighbouring synaptic inputs modulated by the effects of short-term depression and facilitation, $D_k$ and $F_k$ respectively,
\begin{equation}
    \label{Eqtn:synaptic inputs}
    I^{syn}_j = \lambda\sum_{k=1}^8 w_{jk}^{place} x_k D_k F_k
\end{equation}
where $w_{jk}^{place}$ is the weight projecting from place cell $k$ onto place cell $j$. In this model, all these weights are fixed at a value of 1. $\lambda$ takes on a value of 0 or 1 depending on whether MiRo is exploring ($\lambda=0$) or is at the reward ($\lambda=1$). This prevents any synaptic transmissions during exploration, but not whilst MiRo is at the reward (the point at which reverse replays occur). This two-stage approach can be found in similar models as a means to separate an \textit{encoding} stage during exploration from a \textit{retrieval} stage \cite{saravanan2015transition}, and was a key feature of some of the early associative memory models \cite{hopfield1982neural}. Experimental evidence also supports this two-stage process due to the effects of acetylcholine. Acetylcholine levels have been shown to be high during exploration but drop during rest \cite{kametani1990alterations}, whilst acetlycholine itself has the effect of suppressing the recurrent synaptic transmissions in the hippocampal CA3 region \cite{hasselmo1995dynamics}. It is for this reason that the global inhibitory inputs found in \cite{whelan2020fast} are not necessary, as the $\lambda$ term here does functionally the same operation as the inhibitory inputs (inhibition is decreased during reverse replays, thus increasing synaptic transmission), yet is simpler to implement.

$D_k$ and $F_k$ in Equation \ref{Eqtn:synaptic inputs} are respectively the short-term depression and short-term facilitation terms, and for each place cell these are computed as (as in \cite{haga2018recurrent}, but see \cite{tsodyks1998neural,vasilaki2014emergence,esposito2015adaptation}),
\begin{equation}
    \label{eqtn:STD_dynamics}
    \frac{d}{dt}D_k = \frac{1-D_k}{\uptau_{STD}} - x_k D_k F_k
\end{equation}
\begin{equation}
    \label{eqtn:STF_dynamics}
    \frac{d}{dt}F_k = \frac{U-F_k}{\uptau_{STF}} + U \left( 1-F_k \right) x_k
\end{equation}
where $\uptau_{STD}$ and $\uptau_{STF}$ are the time constants, and $U$ is a constant representing the steady-state value for short-term facilitation when there is no neuron activity $(x_k=0)$. $D_k$ and $F_k$ each take on values in the range $[0,1]$. Notice that when $x_k > 0$, short-term depression is driven steadily towards 0, whereas short-term facilitation is driven steadily upwards towards 1. Modification of the time constants allows either short-term depression or short-term facilitation effects to dominate. In this model, the time constants are chosen so that depression is the primary short-term effect. This ensures that during reverse replay events, activity propagating from one neuron to the next quickly dissipates, allowing for stable replays without activity exploding in the network. 


We turn finally  to the intrinsic plasticity term in Equation \ref{eqtn:place_cell_dynamics}, represented by $\psi_j$. Its behaviour, as observed in Equation \ref{eqtn:place_cell_dynamics}, is to scale all incoming synaptic inputs. In \cite{pang2019fast}, Pang and Fairhall used a heuristically developed sigmoid whose output was determined as a function of the neuron's rate. Intrinsic plasticity in their model did not decay once it had been activated. Since our robot often travels across most of the environment, we needed a time decaying form of intrinsic plasticity to avoid potentiating all cells in the network. The simplest form of such time decaying intrinsic plasticity is therefore,
\begin{equation}
    \label{eqtn:intrinsic_plasticity}
    \frac{d}{dt}\psi_j = \frac{\psi_{ss} - \psi_j}{\uptau_{\psi}} + \frac{\psi_{max} - 1}{1 + \text{exp} \left[ -\beta (x_j - x_\psi) \right]}
\end{equation}
with again, $\uptau_{\psi}$ being its time constant, and $\psi_{ss}$ being a constant that determines the steady state value for when the sigmoidal term on the right is 0. All of $\psi_{max}$, $\beta$ and $x_{\psi}$ are constants that determine the shape of the sigmoid. Since $\psi_j$ could potentially grow beyond the value of $\psi_{max}$, we restrict $\psi_j$ so that if $\psi_j > \psi_{max}$, then $\psi_j$ is set to $\psi_{max}$.

In order to initiate a replay event, place cell inputs, computed using Equation (\ref{eqtn:place_cell_input}) with MiRo's current location at the reward, are input into the place cell dynamics (Eqtn \ref{eqtn:place_cell_dynamics}) one second after MiRo reaches the reward, for a duration of 100ms. Intrinsic plasticity for those cells that were most recently active during the trajectory is increased, whilst synaptic conductance in the place cell network is turned on by setting $\lambda=1$. This causes the place cell input to activate only its adjacent cells that were recently active. This effects continues throughout all recently active cells, thus resulting in a reverse replay. Short-term depression ensures that the activity dissipates quickly as it propagates from one neuron to the next.

\subsubsection{Striatal Action Cells}
\label{subsec:Striatal Action Cells}
The action cell values determine how MiRo moves in the environment. All place cells project feedforwardly through a set of plastic synapses to all action cells, as shown in Figure \ref{fig:Miro_with_place_action_cells}B. There are 72 action cells, the value of each drawn from a Gaussian distribution with mean $\tilde{y}_i$ and variance $\sigma^2$,
\begin{equation}
\label{eqtn:action cell computation normal}
    y_i \sim \mathcal{N}\left( \tilde{y}_i, \sigma^2 \right)
\end{equation}
The mean value  $\tilde{y}_i$ is calculated as follows,

\begin{equation}
    \label{eqtn:sigmoid_activation_function}
    \tilde{y}_i = \frac{1}{1 + \text{exp} \left[ -c_1 \sum_{j=1}^{100} w_{ij}^{PC\text{-}AC} x_j - c_2 \right]}
\end{equation}
with $c_1$ and $c_2$ determining the shape of the sigmoid. $w_{ij}^{PC\text{-}AC}$ represents the weight projecting from place cell $j$ onto action cell $i$. The sigmoidal function is one possible choice which results in saturating terms in the RL learning rule (Section \ref{subsubsec:Place Cell to Action Cell Synaptic Plasticity}), an alternative option for instance could have been a linear function.  The action cells are restricted to take on values between 0 and 1, i.e. $y_i \rightarrow [0,1]$, and be interpreted as normalised firing rates. 
 

MiRo moves at a constant forward velocity, whereas the output of the action cells sets a target heading for MiRo to move in. This target heading is allocentric, in that the heading is relative to the arena. The activity for each action cell is denoted as $y_i$ and the target heading as $\theta_{target}$. To find the heading from the action cells, the population vector of the action cell values is computed as follows,
\begin{equation}
    \label{eqtn:theta_target}
    \theta_{target} = \arctan \left( \frac{\sum_i y_i \sin{\theta_i}}{\sum_i y_i \cos{\theta_i}} \right) 
\end{equation}
where $\theta_i$ is the angle coded for by action cell $i$. It is also possible to compute the magnitude of the population vector, which denotes how strongly the action cell activities are promoting a particular heading,
\begin{equation}
    \label{eqtn:M_target}
    m_{target} = \sqrt{\left( \sum_i y_i \sin{\theta_i} \right)^2 + \left( \sum_i y_i \cos{\theta_i} \right)^2}
\end{equation}


For practical reasons, the action cells are computed not only from place cell inputs, but also by a separate module, termed a \textit{semi-random walk} module. This is because the network, particularly in the early stages of exploration when the weights are randomised, is often unable to make useable directional decisions. A simple implementation of a semi-random walk module therefore allows MiRo to explore the environment sensibly, as opposed to erratically when the randomised network weights are used. The details of the \textit{semi random walk} implementation is given below. 

\noindent\textit{\textbf{Semi-Random Walk Module}} --
In cases where the signal provided by the action cells, as computed by Equation \ref{eqtn:M_target} is not strong enough (i.e. less than 1), then MiRo takes a random walk rather than following the direction selected by the action cells. To compute the heading, a small but random value, $\theta_{noise}$, is added to MiRo's current heading,
\begin{equation}
    \theta_{random\_walk} = \theta_{current} + \theta_{noise}
\end{equation}
where $\theta_{noise}$ is a random variable taken from the uniform distribution $\theta_{noise} \sim \text{unif}(-50^\text{o}, 50^\text{o})$. This ensures that MiRo generally keeps moving in its current direction, but is capable of changing slightly to the left or right, though by no more than $50^\text{o}$.

To convert this into action cell values, each action cell is computed as a function of its angular distance from $\theta_{random\_walk}$, in a similar to manner to how the place cell activities were computed as the Cartesian distance of MiRo from the place cell centres,
\begin{equation}
    \label{eqtn:action_cells_from_random_walk}
    y_i^{random\_walk} = y_i^{max} \exp \left[ -\frac{ \left( \theta_{random\_walk} - \theta_i \right)^2}{2\theta_d^2} \right]
\end{equation}
where $y_i^{max}$ determines the maximum value for $y_i$, in this case 1, and $\theta_d$ determines the distribution width, and $\theta_i$ is the angle corresponding to action cell $i$. 

To state this more formally, let the magnitude of the place cell network proposal be (see Equation \ref{eqtn:M_target}),
\begin{equation}
     m_{PC\_proposal} = \sqrt{\left( \sum_i \tilde{y}_i \sin{\theta_i} \right)^2 + \left( \sum_i \tilde{y}_i \cos{\theta_i} \right)^2}
\end{equation}
then the final action cell values are only changed to $y_i = y_i^{random\_walk}$ if $m_{PC\_proposal} < 1$. Else they stay as they are from Equation \ref{eqtn:action cell computation normal}.

\noindent\textbf{\textit{Computing Action Cells During Reverse Replays}} --
The computation for $y_i$ in Equation (\ref{eqtn:action cell computation normal}) is suitable for the exploration stage, but requires a minor modification in order for the action cells to properly replay during reverse replay events. Thus far, $y_i$ is computed either by taking the network's output as determined by the place cell inputs or, if this output is weak, by using a semi-random walk. In order for the $y_i$ term to compute properly in the reverse replay case then, we perform the following,
\begin{equation}
    \label{eqtn:y_i during replay}
    y_i^{replay} = \frac{1}{1 + \text{exp} \left[ -c_1 \sum_{j=1}^{100} \left( w_{ij}^{PC\text{-}AC} + 0.1~ sgn( {e_{ij}^{r}}) \right) x_j - c_2 \right]}
\end{equation}
which is the same computation as Equation (\ref{eqtn:sigmoid_activation_function}), with the only difference being that the place cell to action cell weights have added to them the value 0.1 multiplied by the sign of the eligibility trace for that synapse. The term $e_{ij}^{r}$ represents the value of $e_{ij}$, i.e. a trace of the potential synaptic changes, at the moment of reward retrieval. This effectively stores the history of synaptic activity and adds a transient weight increase to synapses that were recently active. How this eligibility trace is computed is described in Section \ref{subsubsec:Place Cell to Action Cell Synaptic Plasticity}.

Modifying the action cells during replays is necessary so that a reverse replay of the place cells can appropriately reinstate the activity in the action cells \cite{gomperts2015vta}. Without this change the reverse replays would offer no additional benefits. This modification acts like a synaptic tag that activates at reward retrieval only and provides temporary synaptic modifications, according to the sign of the eligibility trace, during the reverse replay stage. Despite making this assumption, this temporary change in synaptic strengths is similar in nature to that of acetylcholine levels modifying synaptic conductances during replay events in the hippocampus \cite{hasselmo1995dynamics}. In other words, synaptic weights (and their modifications) are suppressed during exploration, but are manifest during the replay stage.


We also tested the rule using a weaker assumption, adding only a value of 0.1 for any synapse in which $e_{ij} > 0$, whilst adding nothing for synapses where $e_{ij} < 0$. However, this was shown to perform worse than even the non-replay case. On closer inspection, the reason for this proved to be the loss of causality during a replay event. Since replays activate multiple cells simultaneously, those synapses that may have had $e_{ij} < 0$ will be influenced by neighbouring place and action cell activities without having the negative eligibility to counter those effects. This influence caused them to increase their weights, as opposed to decreasing, which would be the proper direction given a negative value for $e_{ij}$.

\subsubsection{Place Cell to Action Cell Synaptic Plasticity}
The learning rule has been derived using a policy gradient reinforcement learning method \cite{sutton2018reinforcement}. Its form is that of a three factor learning rule with an eligibility trace \cite{gerstner2018eligibility}. The full derivation for the learning rule is presented in the Appendix.

When MiRo is exploring, a learning rule of the following is implemented
\begin{equation}
    \label{eqtn:learning_rule_main_text}
    \frac{dw_{ij}^{PC\text{-}AC}}{dt} = \frac{\eta }{\sigma^2} R e_{ij}
\end{equation}
where $R$ is a reward value, whilst the term $e_{ij}$ represents the eligibility trace and is a time decaying function of the potential weight changes, determined by,
\begin{equation}
    \label{eqtn:eligibility_trace_main_text}
    \frac{de_{ij}}{dt} = -\frac{e_{ij}}{\uptau_e} + \left(y_i - \tilde{y}_i\right) \left( 1-\tilde{y}_i \right) \tilde{y}_i x_j.
\end{equation}

During reverse replays however, the activity of the action cells are given by $y_i^{replay}$. In order to retain mathematically accuracy in the derivation (see Appendix), we derived a similar learning rule from a supervised learning approach having the following form

\begin{equation}
    \label{eqtn:learning_rule_replay_main_text}
    \frac{dw_{ij}^{PC\text{-}AC}}{dt} = \eta^\prime e_{ij},
\end{equation}
where the eligibility trace is determined by,
\begin{equation}
    \label{eqtn:eligibility_trace_replay_main_text}
    \frac{de_{ij}}{dt} = -\frac{e_{ij}}{\uptau_e} + \left(y_i^{replay} - \tilde{y}_i\right) \left( 1-\tilde{y}_i \right) \tilde{y}_i x_j,
\end{equation}
We have set $\eta' = \eta / \sigma^2$ and let $R = 1$ at the reward location in our simulations, which renders the RL rule and the supervised learning rule equivalent.

\subsubsection{Population weight vector for a single place cell}

The population weight vector for a single place cell is computed as,
\begin{equation}
\label{eqtn:population vector}
    (w_j^x, w_j^y) = \left( \sum_{i=1}^{72} w_{ij}^{PC\text{-}AC} \cos{\theta_i}\, , \, \sum_{i=1}^{72} w_{ij}^{PC\text{-}AC} \sin{\theta_i} \right)
\end{equation}
where $(w_j^x, w_j^y)$ represents the $x$ and $y$ components for the weight population vector of the $j^{th}$ place cell, $w_{ij}^{PC\text{-}AC}$ is the value of the weight from place cell $j$ onto action cell $i$, and $\theta_i$ is the heading direction that action cell $i$ codes for. The magnitude of the population weight vector can then be computed as,
\begin{equation}
\label{eqtn:population vector magnitude}
    M_{w_j} = \sqrt{\left(w_j^x\right)^2 + \left(w_j^y\right)^2}
\end{equation}
The population weight vector depicts the preferred direction of MiRo when placed at the center of the location of the place cell. 

\subsubsection{Implementation}
A description of the full implementation process is provided here, with an overview of the algorithmic implementation presented in the Supplementary Material.

\noindent\textbf{\textit{Initialisation}} -- At the start of a new experiment, the weights that connect the place cells to the action cells are randomised and then normalised,
\begin{equation}
    \label{eqtn:normalise_weights}
    w_{ij}^{PC\text{-}AC} \leftarrow \frac{w_{ij}^{PC\text{-}AC}}{\sum_i w_{ij}^{PC\text{-}AC}}.
\end{equation}
All the variables for the place cells are set to their steady state conditions for when no place specific inputs are present, and the action cells are all set to zero. MiRo is then placed into a random location in the arena.

\noindent\textbf{\textit{Taking Actions}} -- There are three main actions MiRo can make, depending on whether the reward it receives is positive ($R=1$) and is therefore at the goal, negative ($R=-1$) such that MiRo has reached a wall, or zero ($R=0$) for all other cases. If the reward is 0, the action cell values, $y_i$ is computed according to Equation \ref{eqtn:action cell computation normal}, or $y_i^{random\_walk}$ is computed from Equation \ref{eqtn:action_cells_from_random_walk} if $m_{PC_proposal} < 1$, letting then $y_i = y_i^{random\_walk}$. From this, a heading is computed using Equation \ref{eqtn:theta_target}. MiRo moves at a constant forward velocity with this heading, with a new heading being computed every 0.5s. If MiRo reaches a wall, a wall avoidance procedure is used, turning MiRo round 180$^{\text{o}}$. Finally, if MiRo reaches the goal, it pauses there for 2s, after which it heads to a new random starting location.

\noindent\textbf{\textit{Determining Reward Values}} -- As described above, there are three reward values that MiRo can collect. If MiRo has reached a wall, a reward of $R = -1$ is presented to MiRo for a period of 0.5s, which tends to occur during MiRo's wall avoidance procedure. If MiRo has found the goal, a reward of $R = +1$ is presented to it for a period of 2s. And if neither of these conditions are true, then MiRo receives no reward, i.e. R = 0.

\noindent\textbf{\textit{Initiating Reverse Replays}} -- Reverse replays are only inititiated when MiRo reaches the goal location, but not for when MiRo is avoiding a wall. For the case in which reverse replays are initiated, $\lambda$ is set to 1 to allow hippocampal synaptic conductance, and the place specific input for MiRo's position whilst at the goal, $I_j^{place}$, is injected 1s after MiRo first reaches the goal for a total time of 100ms. With synaptic conductance enabled, and due to intrinsic plasticity, this initiates reverse replay events initiating at the goal location and traveling back through the recent trajectory in the place cell network. An example of a reverse replay can be found in the Supplementary Material. Whilst learning is done as standard in the non-replay stage using Equations \ref{eqtn:learning_rule_main_text} and \ref{eqtn:eligibility_trace_main_text} when MiRo first reaches the goal, once the replays start learning is done using the supervised learning rule of Equations \ref{eqtn:learning_rule_replay_main_text} and \ref{eqtn:eligibility_trace_replay_main_text}.

\noindent\textbf{\textit{Updating Network Variables}} -- Regardless of whether MiRo is exploring, avoiding a wall, or is at the goal and is initiating replays, all the network variables, including the weight updates, occur for every time step of the simulation. It is only when MiRo has reached the goal, gone through the 2s of reward collection, and is making its way to a new random start location that all the variables are reset as in the Initialisation step above (though excluding the randomisation of the weights). This would then begin a new trial in the experiment.

\subsubsection{Model parameter values}

All parameter values used in the Hippocampal network are summarised in Table \ref{tab:model parameters hippocampal model}, and those for the Striatal network in Table \ref{tab:action cell parameter values}. Values for $\eta$ and $\uptau_e$ are specified appropriately in the results, since they are modified across various experiments.

\begin{table}[h!]
\begin{center}
 \begin{tabular}{||c c||}
 \hline
 Parameter & Value \\ [0.5ex] 
 \hline\hline
 $\alpha$ & $1C^{-1}$ \\ 
 \hline
 $\epsilon$ & $2A$ \\
 \hline
 $\uptau_I$ & $0.05s$ \\
 \hline
 $I^p_{max}$ & $50A$ \\
 \hline
 $d$ & $0.1m$ \\
  \hline
 $\lambda$ & $0$ or $1$, see text \\ 
 \hline
 $\uptau_{STD}$ & $1.5s$ \\
 \hline
 $\uptau_{STF}$ & $1s$ \\
 \hline
 $U$ & 0.6 \\
 \hline
 $\psi_{ss}$ & 0.1 \\
 \hline
 $\psi_{max}$ & 4 \\
 \hline
 $\uptau_{\psi}$ & $10s$ \\
 \hline
 $\beta$ & $1$ \\
 \hline
 $x_{\psi}$ & $10Hz$ \\
 \hline
\end{tabular}
\caption{Summarising the model parameter values for the hippocampal network used in the experiments. All these parameters are kept constant across all experiments.}
\label{tab:model parameters hippocampal model}
\end{center}
\end{table}

\begin{table}
\begin{center}
 \begin{tabular}{||c c||}
 \hline
 Parameter & Value \\ [0.5ex] 
 \hline\hline
 $c_1$ & 0.1 \\ 
 \hline
 $c_2$ & 20 \\
 \hline
 $\sigma$ & 0.1 \\
 \hline
 $\theta_d$ & 10 \\
 \hline
 $\uptau_e$ & \textit{See text} \\
 \hline
 $\eta$ & \textit{See text} \\ 
 \hline
\end{tabular}
\caption{Summarising the model parameter values for the striatal network used in the experiments. Except for the learning rate, $\eta$, and the eligibility trace time constant, $\uptau_e$, all other parameters are kept constant for all experiments.}
\label{tab:action cell parameter values}
\end{center}
\end{table}

\newpage
\section{Results}

This results section is divided into two subsections. Presented first are the results for when running the model without reverse replays, to demonstrate the functionality of the network and the learning rule. Following this, the model is then run with reverse replays, with these results being compared to the non-replay case. All model parameters and the learning rule are kept equal between the two cases to facilitate the comparison, however when we compare the two models in terms of performance we optimise the key parameters deferentially for each model, comparing best with best performance.


\subsection{Learning Rule Without Reverse Replays}

\begin{figure}[t!]
    \centering
    \includegraphics[width=0.6\linewidth]{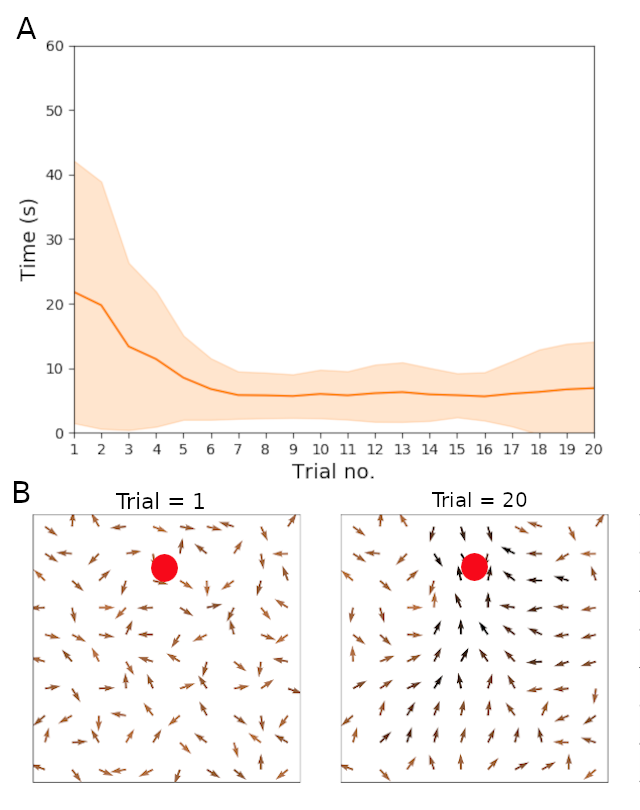}
    \caption{Results for the non-replay case in order to test that the derived learning rule performs well. Parameters used were $\eta = 0.01$ and $\uptau_e = 1s$. A) Plot showing the average time to reach goal (red line) and standard deviations (shaded area) over 20 trials. Averages and standard deviations are computed from 20 independent experiments. B) Weight population vectors at the start of trial 1 versus at the end of trial 20 in an example experiment. Magnitudes for the vectors are represented as a shade of colour; the darker the shade, the larger the magnitude. Red dots indicate the goal location.}
    \label{fig:Non_replay_case}
\end{figure}

We first demonstrate the functionality of the learning rule (Equations \ref{eqtn:learning_rule_main_text} and \ref{eqtn:eligibility_trace_main_text}), without reverse replays. Figure \ref{fig:Non_replay_case}A shows the results for the time taken to reach the hidden goal as a function of trial number, averaged across 20 independent experiments. The time to reach the goal approaches the asymptotic performance after around 5 trials. Note however that there appears to be larger variance towards the final two trials. Further trials were later run in order to test whether this increased variability in performance was significant or not (see Section \ref{subsubsec:Comparison of Best Cases}). Figure \ref{fig:Non_replay_case}B displays the weight vector for the weights projecting from the place cells to the action cells. We note that after 20 trials the arrows in general point towards the direction of the goal.

\subsection{Effect of Reverse Replays on Performance}
\label{subsection:Effect of Reverse Replays on Performance}
We then ran the model with reverse replays, implementing the learning rule of Equations \ref{eqtn:learning_rule_replay_main_text} and \ref{eqtn:eligibility_trace_replay_main_text}, using first the same learning rate and eligibility trace time constant as in the non-replay case above. The performance average showed not to have any significant difference ($p>0.05$ across 18 trials in a Wilcoxon Signed-Rank Test). Average time to reach goal over the last 10 trials is 6.21s in the non-replay case and 6.92s in the replay case (data not shown, see Supplementary Material). This suggests in the first instance that replays are at least as good when compared to the best case non-replay, which was also confirmed when comparing individually optimised parameters (learning rate and eligibility time constant) for each network. Further results on performance of varying the learning rate and eligibility trace time constant are presented next.

\subsubsection{Reducing the Eligibility Trace Time Constant}
\begin{figure}[t!]
    \centering
    \includegraphics[width=0.8\textwidth]{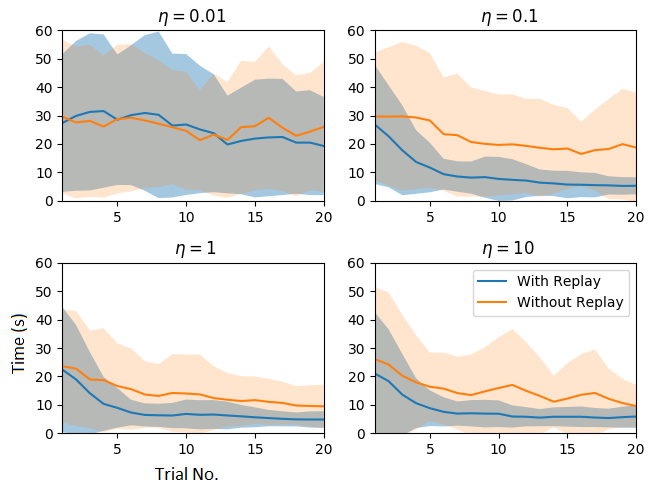}
    \caption[Comparing the effects of a small eligibility trace time constant with and without reverse replays]{Comparing the effects of a small eligibility trace time constant with and without reverse replays. $\uptau_e=0.04s$ across all figures. Thick lines are averages across 40 independent experiments, with shaded areas representing one standard deviation. Moving averages, averaged across 3 trials, are plotted here for smoothness.}
    \label{fig:tau=0.04s}
\end{figure}

Given the standard, non-replay model requires the recent history to be stored in the eligibility trace, it follows that having too small an eligibility trace time constant might negatively impact the performance of the model. This is due to the time constant reflecting how far back the information about the Reward will be ``transmitted''. Reverse replays however have the potential to compensate for this, since the recent history is also stored, and then replayed, in the place cell network. Figure \ref{fig:tau=0.04s} shows the effects on performance of significantly reducing the eligibility trace time constant (to $\uptau=0.04s$). Both cases, with and without reverse replays, are compared. If the learning rate is too small ($\eta = 0.01$) then for neither case is there any learning. But as the learning rate is increased, having reverse replays shows to significantly improve performance. Similar results are found for a larger, but still small, eligibility trace time constant of $\uptau_e = 0.2s$ (see Supplementary Material).


\subsubsection{Comparing Differences in Synaptic Weight Changes}

\begin{figure}[t!]
    \centering
    \includegraphics[width=0.76\textwidth]{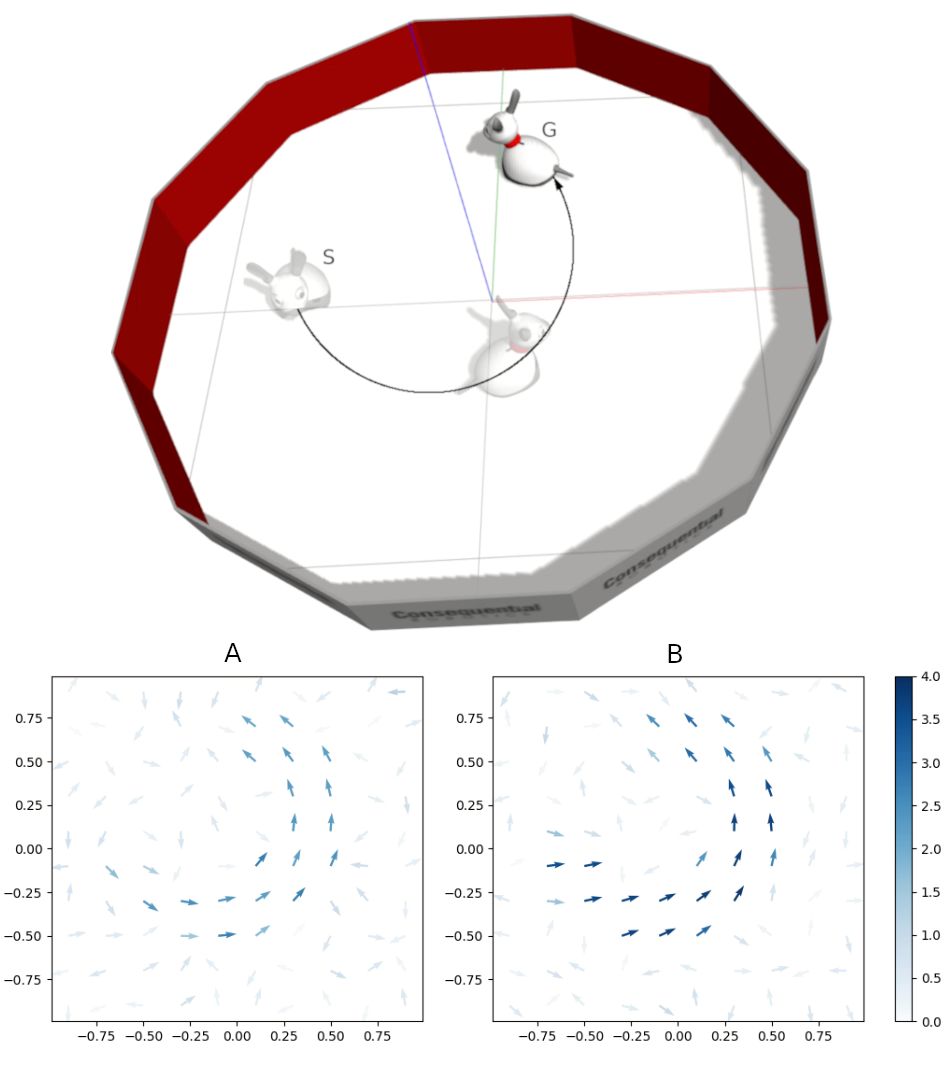}
    \caption[Population weight vectors after reward retrieval in the non-replay and replay cases]{Population weight vectors after reward retrieval in the non-replay and replay cases. Top figure shows the path taken by MiRo, where S represents the starting location and G the goal location. Plots show weight population vectors for the non-replay case (A) and standard replay case (B) with $\uptau_e = 1s$; $\eta = 0.1$. The color scale represents the magnitudes for each weight vector.}
    \label{fig:weight_changes_replay_vs_nonreplay}
\end{figure}

An interesting comparison can be shown between the magnitudes of weight changes for the replay case and non-replay cases. Figure \ref{fig:weight_changes_replay_vs_nonreplay} shows the population vectors of the weights after reward retrieval. Population vectors for the weights are computed according to Equations \ref{eqtn:population vector} and \ref{eqtn:population vector magnitude}. There are two observations to be made here. First is that the weight magnitudes are greater when reverse replays are implemented, which is expected since activity replay offers additional information to the synaptic changes. And second is that the direction of the population weight vectors themselves are slightly different, particularly in the location at the start of the trajectory. In particular, the weight vectors point more towards the goal location in the replay case, whereas the non-replay case has weight vectors pointing along the direction of the path taken by the robot. Whilst only one case has been depicted here, this is representative of a number of cases for various parameter values. 

\subsubsection{Performance Across Parameter Space}
\begin{figure}[t!]
    \centering
    \includegraphics[width=0.9\textwidth]{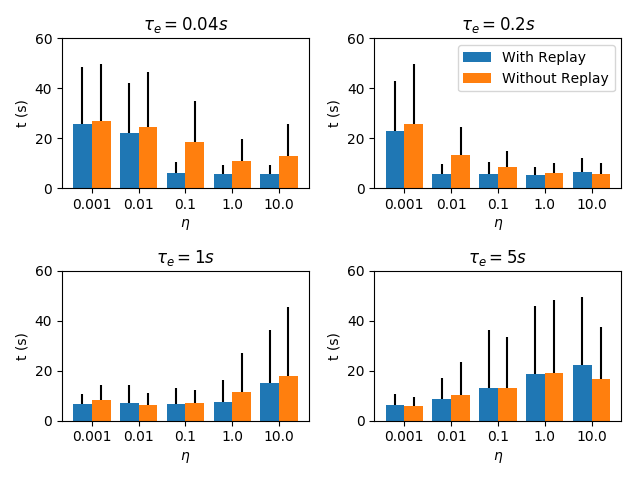}
    \caption[Comparing average performance across the last 10 trials]{Comparing average performance across a range of values for $\uptau_e$ and $\eta$. Bars show the average time taken to reach the goal. These plots are found by first averaging across 40 independent experiments (as shown in Figure \ref{fig:tau=0.04s} for instance), and then averaging over the last 10 trials. Error bars indicate the 10 trial average of standard deviations.}
    \label{fig:comparison across parameter space}
\end{figure}

We investigated the robustness of the performance across various values of $\uptau_e$ and $\eta$. Figure \ref{fig:comparison across parameter space} displays the average performance over the last 10 trials, comparing again with replays versus without replays. There are perhaps two noticeable observations to make here. Firstly, when the eligibility trace time constant is small, employing reverse replays shows considerable improvements in performance over the non-replay case across the various values of learning rates. Learning still exists in the non-replay case, however, it is noticeably diminished compared with the replay case. Secondly, although this marked improvement in performance vanishes for larger eligibility trace time constants, reverse replays do not at the very least hinder performance.

\subsubsection{Comparison of Best Cases}
\label{subsubsec:Comparison of Best Cases}
\begin{figure}[b!]
    \centering
    \includegraphics[width=0.8\textwidth]{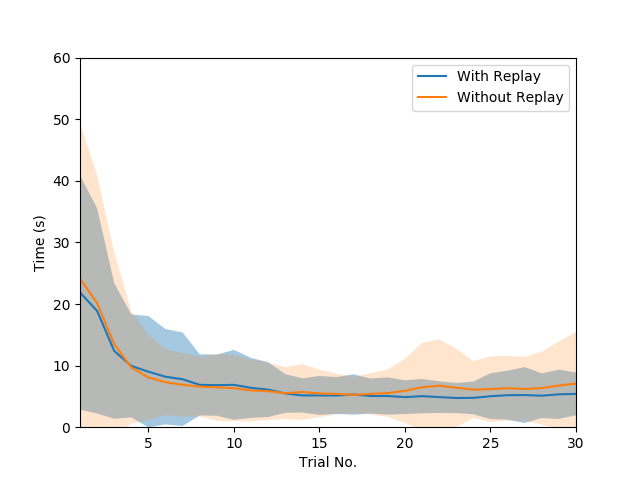}
    \caption[Comparing the best cases with and without reverse replays]{Comparing the best cases with and without reverse replays. With reverse replays the parameters are $\uptau_e = 0.04s$, $\eta = 1$. Without reverse replays the parameters are $\uptau_e = 1s$, $\eta = 0.01$. The means (solid lines) and standard deviations (shaded regions) are computed across 40 independent experiments.}
    \label{fig:comparing_best_cases}
\end{figure}

Figure \ref{fig:comparing_best_cases} compares the results for the best cases with and without reverse replays. To achieve these results we optimised $\uptau_e$ and $\eta$ independently for each case and run a total of 30 trials. The reason for this was a suspected instability in the non-replay case when the amount of trials increased as indicated in Figure \ref{fig:Non_replay_case}. A Wilcoxin signed-rank test was run on all trials for the two cases, and for 8 of the last 12 trials (trials 19-30), there were significant differences between the two ($p<0.05$, full table of results can be found in the Supplementary material) despite there being no significant differences in trials 0-18 (Section \ref{subsection:Effect of Reverse Replays on Performance} above).

This instability in the non-replay case is not observed in the case with replays. We also note the difference in parameters for the best cases. With reverse replays the parameters are $\uptau_e = 0.04s$, $\eta = 1$, whereas without reverse replays they are $\uptau_e = 1s$, $\eta = 0.01$. We speculate that the necessary choice in the eligibility time constant for the non-replay case (i.e. that it necessarily needs to be large enough to store the trajectory history) is the cause of this instability. 

\section{Discussion}
\label{sec:Discussing the Model and Results}
Hippocampal reverse replay has long been implicated in reinforcement learning \cite{foster2006reverse}, but how the dynamics of hippocampal replay produce behavioural changes, and why hippocampal replay could be important in learning, are ongoing questions. By embodying first a hippocampal-striatal inspired model \cite{vasilaki2009spike} into a simulated MiRo robot, and then augmenting it with a model of hippocampal reverse replay \cite{whelan2020fast}, we have been able to examine the link between hippocampal replay and behavioural changes in a spatial navigation task. We have shown that reverse replays can enable quicker reinforced learning, as well as generating more robust behavioural trajectories over repeated trials.

In the three-factor, synaptic eligibility trace hypothesis, the time constants for the traces have been argued to be on the order of a few seconds, necessary for learning over behavioural time scales \cite{gerstner2018eligibility}. However, results here indicate that due to reverse replays, it is not necessary for synaptic eligibility trace time constants to be on the order of seconds -- a few milliseconds is sufficient. The synaptic eligibility trace is still required here for storing the history; it just does not matter how much of the eligibility trace is stored, it is only important that enough is stored for effective reinstatement during a reverse replay. It has also been argued that neuronal, as opposed to synaptic, eligibility traces could be sufficient for storing a memory trace, as in the two-compartmental neuron model of \cite{brea2016prospective}. Intrinsic plasticity in this model is not unlike a neuronal eligibility trace, storing the memory trace within the place cells for reinstatement at the end of a rewarding episode.

It could be the case that reverse replays speed up learning by introducing an additional source of information regarding past states, and the results shown here provide some support for this. Experimental evidence does show for instance that disruption of hippocampal ripples during awake states, when reverse replays occur, does disrupt but not completely diminish spatial learning in rats \cite{jadhav2012awake}. Whilst the longer eligibility trace time constants in this model ($\uptau_e= 1s, 5s$) do not show diminished performance without reverse replays, the smaller time constants ($\uptau_e = 0.04s, 0.2s$) do. Hence, these results support the view that reverse replays enhance, rather than provide entirely, the mechanism for learning. Beyond reverse replays however, forward replays have been known to occur on multiple occasions for up to 10 hours post-exploration \cite{giri2019hippocampal}, which could be more important for memory consolidation than awake reverse replays \cite{girardeau2009selective,ego2010disruption}.

In the best case comparison (Figure \ref{fig:comparing_best_cases}), it is clear why a sufficiently large, yet not overly large, eligibility trace time constant for the non-replay case gives best performance -- it must store a suitable amount of the trajectory history for learning. If the eligibility trace time constant were too small, it would not store enough of the history, whereas too large and it stores sub-optimal or unnecessary trajectories that go too far back in time. Yet the non-replay model became more unstable as the number of trials increased, as shown in Figure \ref{fig:comparing_best_cases}. One explanation for this is that the eligibility trace time constant necessary for learning in non-replay had to be large enough to store trajectory histories, but doing this increases the probability that sub-optimal paths may be learned. For the replay case however, since the trajectory was replayed during learning, it was not necessary to have such a large eligibility trace time constant. Sub-optimal paths going further back in time are therefore no longer as strongly learned. Furthermore, replays are able to modify slightly the behavioural trajectories. By looking at the effects in the weight vectors of Figure \ref{fig:weight_changes_replay_vs_nonreplay}, it is apparent that the weight vectors closer to the start location are shifted to point more towards the goal in the replay case. Reverse replays could help in solving the exploration-exploitation problem in RL \cite{sutton2018reinforcement}, since they could simulate more optimal trajectories that were not explicitly taken during behaviour.

In this model, there are two sets of competing behaviours during the exploratory stage -- the memory guided behaviour of the hippocampus and the semi-random walk behaviour -- which are heuristically selected for based on the signal strength of the hippocampal output: If the hippocampal output does not express strongly for a particular action, the semi-random walk behaviour is implemented instead. An interesting comparison with the basal ganglia, and its input structure the striatum, could be made here, since these structures are thought to play a role in action selection \cite{mink1996basal,grillner2005mechanisms,prescott2006robot,redgrave2017phasic}. A basic interpretation of this action selection mechanism is that the basal ganglia receives a variety of candidate motor behaviours, each of which are perhaps mutually incompatible, but from which the basal ganglia must select one (or more) of these behaviours for expressing \cite{gurney2001computational1,gurney2001computational2}. Since the selection of an action in our model is determined from the striatal action cell outputs, it appears likely that this selection would occur within the basal ganglia.

But perhaps more interesting is that in the synaptic learning rule presented here, the difference between the action selected, $y_i$, and the hippocampal output, $\tilde{y}_i$,
is used to update synaptic strengths. One interpretation for this could be that this difference behaves as an error signal, signalling to the hippocampal-striatal synapses how ``good'', or how ``close'', their predictions were in generating behaviours that led towards rewards. But how might this be implemented in the basal ganglia? Whilst the striatum acts as the input structure to the basal ganglia, neuroanatomical evidence shows that the basal ganglia sub-regions loop back on one another \cite{gurney2001computational1}, and that in particular the striatum sends inhibitory signals to the substantia nigra (SN), which in turn projects back both excitatory and inhibitory signals via dopamine (D1 and D2 receptors respectively) to the striatum \cite{gerfen1990d1,harsing1997influence}. There is therefore a potential mechanism for appropriate feedback to the hippocampal-striatal synapses in order to provide this error signalling, and an exploration of this error signal hypothesis could be a potentially interesting research endeavour.

\section{Conclusion}
This work has explored the role that reverse replays may have in biological reinforcement learning. As a baseline, we have derived a  
policy gradient Reinforcement Learning rule, which we employed to associate actions with place cell activities. This is a three factor learning rule with an eligibility trace, where the eligibility trace stores the pairwise co-activities of place and action cells. The learning rule was shown to perform successfully when applied to a simulated MiRo robot for a Morris water-maze like task. We further augmented the network and learning rule with reverse replays, which acted to reinstate recent place and action cell activities. The effect of these replays was that learning was significantly improved for circumstances in which eligibility traces did not store sufficient activity history. In addition, this had the effect of generating more stable performance as the number of trials increased. Learning with reverse replays was improved upon the case without replays, yet learning was still achievable without replays. Reverse replay may therefore enhance reinforcement learning in the hippocampal-striatal network whilst not necessarily providing its core mechanism.





\section*{Funding}
This work has been in part funded by the EU Horizon 2020 programme through the FET Flagship Human Brain Project (HBP-SGA2: 785907; HBP-SGA3: 945539).

\section*{Acknowledgments}
The authors would like to thank Andy Philippides and Michael Mangan for their valuable input and useful discussions.

\bibliography{references}

\bibliographystyle{plain}

\clearpage
\section*{Appendix - Mathematical Derivation of the Place-Action Cell Synaptic Learning Rule}
\noindent\textit{\textbf{Derivation of the reinforcement learning rule}} -- We derive a policy gradient rule \cite{sutton2018reinforcement} following  \cite{vasilaki2009spike}, but here we use continuous valued neurons instead of spiking neurons. The expectation for the rewards earned in an episode of duration $T$ is given by,
\begin{equation}
    \label{eqtn:average_reward}
    \left<R\right>_T = \int_X \int_Y \ R
    \left({\bf x}, {\bf y}\right)P_w\left({\bf x},{\bf y}\right) d{\bf y}d{\bf x}
\end{equation}
where $X$ is the space of the inputs of and $Y$ the space of the output of the network, and $P_w\left({\bf x},{\bf y}\right)$ the probability that the network has input ${\bf x}$ and output ${\bf y}$, parametrised by the weights. 

We can decompose the probability, $P_w\left({\bf x},{\bf y}\right)$ (see decomposition of the probability in \cite{vasilaki2009spike}) as,
\begin{equation}
    \label{eqtn:probability_distribution}
    P_w\left({\bf x},{\bf y}\right) = \prod_{j} g_j\left({\bf x},{\bf y}\right) \prod_i h_i\left({\bf x},{\bf y}\right),
\end{equation}
where $h_i$ is the probability the $i$-th action cell generates output ${\bf y}_j$ contained in  ${\bf y}$ , when the network receives input ${\bf x}$. Similarly  $g_j$ is the probability for the activity produced by the $j$-th place cell given its input. We then wish to calculate the partial derivative over a weight $w_{kl}$ of the expected reward,

\begin{equation}
    \label{eqtn:differentation of average reward}
    \frac{\partial \left<R_T\right>}{\partial w_{kl}} = \int_X \int_Y \ R
    \left({\bf x}, {\bf y}\right) \frac{\partial P_w\left({\bf x},{\bf y}\right)}{\partial w_{kl}}~d{\bf y}d{\bf x} .
\end{equation}
To do so,  we take into account that $P_w\left({\bf x},{\bf y}\right)= \left[ \frac{P_w\left({\bf x},{\bf y}\right)}{h_k\left({\bf x},{\bf y} \right)} \right]h_k\left({\bf x},{\bf y}\right)$, where the term in square brackets does not depend on $w_{kl}$ since we remove its contribution from $P_w\left({\bf x},{\bf y}\right)$ by dividing with $h_k\left({\bf x},{\bf y}\right)$. We can then write,

\begin{equation}
    \label{eqtn:trick}
    \frac{\partial P_w\left({\bf x},{\bf y}\right)}{\partial w_{kl}}= P_w\left({\bf x},{\bf y}\right)\frac{\partial~ log~h_k\left({\bf x},{\bf y}\right)}{\partial w_{kl}}.
\end{equation}
This leads to,
\begin{equation}
    \label{eqtn:differentation of average reward2}
    \frac{\partial \left<R_T\right>}{\partial w_{kl}} = \int_X \int_Y \ R
    \left({\bf x}, {\bf y}\right) P_w\left({\bf x},{\bf y}\right)\frac{\partial~ log~h_k\left({\bf x},{\bf y}\right)}{\partial w_{kl}}~d{\bf y}d{\bf x} .
\end{equation}
To proceed, we need to consider the distribution of the activities of the action cells $h_k$. This we choose to be a Gaussian function with mean $\bar{y}_k$  and variance $\sigma^2$ (see also section ``Striatal Action Cells''),
\begin{equation}
    \label{eqtn:probability_distribution_of_action_cells}
    h_k\left(X,Y\right) = \frac{1}{\sigma \sqrt{2\pi}}\exp{\left(-\frac{\left(y_k - \tilde{y}_k\right)^2}{2\sigma^2}\right).}
\end{equation}
The mean of the distribution is calculated by $\tilde{y}_k = f_s\left( c_1 \sum_j w_{kj} x_j +c_2 \right)$, see also Equation \ref{eqtn:sigmoid_activation_function},  where $f_s$ is a sigmoidal function. We note that a different choice of function would have resulted in a variant of this rule. 
Therefore, 
\begin{equation}
    \label{eqtn:differentation of average reward3}
  \frac{\partial~ log~h_k\left({\bf x},{\bf y}\right)}{\partial w_{kl}} = c_1 \frac{y_k-\tilde{y}_k}{\sigma^2} \left( 1 - \tilde{y}_k \right) \tilde{y}_k  x_l.
\end{equation}
Replacing \ref{eqtn:differentation of average reward3} in \ref{eqtn:differentation of average reward2} we end up with,

\begin{equation}
    \label{eqtn:differentation of average reward4}
    \frac{\partial \left<R_T\right>}{\partial w_{kl}} = \int_X \int_Y c_1 \ R
    \left({\bf x}, {\bf y}\right) P_w\left({\bf x},{\bf y}\right) \frac{y_k-\tilde{y}_k}{\sigma^2} \left( 1 - \tilde{y}_k \right) \tilde{y}_k x_l ~d{\bf y}d{\bf x} .
\end{equation}
Then the batch update rule is given by,
\begin{equation}
    \label{eqtn:differentation of average reward5}
    \frac{dw_{kl}}{dt}= \eta \int_X \int_Y \ R
    \left({\bf x}, {\bf y}\right) P_w\left({\bf x},{\bf y}\right) \frac{y_k-\tilde{y}_k}{\sigma^2} \left( 1 - \tilde{y}_k \right) \tilde{y}_k x_l ~d{\bf y}d{\bf x} .
\end{equation}
The batch rule indicates that we need to average the term $ R
    \left({\bf x}, {\bf y}\right) \frac{y_k-\tilde{y}_k}{\sigma^2} \left( 1 - \tilde{y}_k \right) \tilde{y}_k  x_l$ across many trials. When an on-line setting is considered, the average is naturally rising from sampling throughout the episodes. Hence the on-line version of this rule is given by,
\begin{equation}
    \label{eqtn:differentation of average reward5}
     \frac{dw_{kl}}{dt}= \eta~R
    \left({\bf x}, {\bf y}\right)~\frac{y_k-\tilde{y}_k}{\sigma^2}  \left( 1 - \tilde{y}_k \right) \tilde{y}_k x_l ,
\end{equation}
with the factor $c_1$ absorbed in the learning rate.
We note however that this rule is appropriate for scenarios where reward is immediate. To deal with cases of distant rewards, such as ours where reward comes at the end of a sequence of actions, we need to resort to eligibility traces. Our rule is similar to REINFORCE with multiparameter distribution \cite{williams1992simple}; we differ by having a continuous time formulation and a different parametrisation of the neuronal probability density function. Further, in our case we do not learn the variance of the probability density function. 


\label{subsubsec:Place Cell to Action Cell Synaptic Plasticity}
We introduce an eligibility trace by updating the weights connecting the place cells to the action cells, $W^{PC\text{-}AC}$ by
\begin{equation}
    \label{eqtn:learning_rule}
    \frac{dw_{ij}^{PC\text{-}AC}}{dt} = \frac{\eta }{\sigma^2} R
    \left({\bf x}, {\bf y}\right)  e_{ij}
\end{equation}
The term $e_{ij}$ represents the eligibility trace, see also \cite{sutton2018reinforcement}, and is a time decaying function of the potential weight changes, determined by,
\begin{equation}
    \label{eqtn:eligibility_trace}
    \frac{de_{ij}}{dt} = -\frac{e_{ij}}{\uptau_e} + \left(y_i - \tilde{y}_i\right) \left( 1-\tilde{y}_i \right) \tilde{y}_i x_j.
\end{equation}

\noindent\textit{\textbf{Derivation of the supervised learning rule}} -- During replays, we assume that synapses between place and action cells change to minimise the function

\begin{equation}
    \label{eqtn:sup_error}
    E=\frac{1}{2}\sum_i\left(y_i^{replay} - \tilde{y}_i\right)^2,
\end{equation}
in other words, we assume that during the replay Equation \ref{eqtn:y_i during replay} provides a fixed target value for the mean of the Gaussian distribution of the action cells at time $t$. In what follows we consider the target constant for the shape of the derivation and consistency with the form of the reinforcement learning rule, but in fact this target changes as time and consequently the weights from place to action cells change, making the rule unstable, but stabilising under a short, fixed length of reply time. 
Taking the gradient over the error function with respect to the weight $w_{kl}$, when considering the ``target'' activity for the action cells fixed, leads to the backpropagation update rule for a single layer network 

\begin{equation}
    \label{eqtn:sup_rule}
     \frac{dw_{kl}}{dt}= \eta^\prime~\left(y_k^{replay}-\tilde{y}_k\right) \tilde{y}_k \left( 1 - \tilde{y}_k \right) x_l,
\end{equation}
where $\eta^\prime$ is the learning rule, in our simulations $\eta^\prime=\eta/\sigma^2$ similar to the reinforcement learning rule. Also for consistency with the reinforcement learning rule formulation, we introduce an eligibility trace by updating the weights connecting the place cells to the action cells, $W^{PC\text{-}AC}$ by
\begin{equation}
    \label{eqtn:learning_rule_replay}
    \frac{dw_{ij}^{PC\text{-}AC}}{dt} = \eta^\prime e_{ij},
\end{equation}
where the eligibility trace is determined by,
\begin{equation}
    \label{eqtn:eligibility_trace_replay}
    \frac{de_{ij}}{dt} = -\frac{e_{ij}}{\uptau_e} + \left(y_i^{replay} - \tilde{y}_i\right) \left( 1-\tilde{y}_i \right) \tilde{y}_i x_j,
\end{equation}
where again the time constant $\tau_e$ is the same as in the reinforcement learning rule.

In the case of replays then, when the robot has reached its target, it first learns using the standard learning rule as in Equations \ref{eqtn:learning_rule} and \ref{eqtn:eligibility_trace}. After 1s, a replay event is initiated, and learning is then done using the supervised learning rule here, using Equations \ref{eqtn:learning_rule_replay} and \ref{eqtn:eligibility_trace_replay}. By setting the reward value to $R=1$, we can ensure that both the RL learning rule and the supervised learning rule are equal.

\end{document}